\documentclass[%
reprint,
superscriptaddress,
 amsmath,amssymb,
 aps,
]{revtex4-2}

\usepackage{graphicx}
\usepackage{dcolumn}
\usepackage{bm}
\usepackage{xcolor}
\graphicspath{{Figures/}}
\usepackage{gensymb}

\begin{document}

\preprint{}

\title{Pair-loaded electron-only magnetic reconnection using laser-driven capacitor coils}

\author{Brandon K. Russell}%
 \email{br2114@princeton.edu}
\affiliation{Department of Astrophysical Sciences, Princeton University, Princeton, New Jersey 08544, USA}

\author{Qian Qian}
\affiliation{G\'{e}rard Mourou Center for Ultrafast Optical Science, University of Michigan, 2200 Bonisteel Boulevard, Ann Arbor, Michigan 48109, USA}

\author{Rebecca Fitzgarrald}
\affiliation{G\'{e}rard Mourou Center for Ultrafast Optical Science, University of Michigan, 2200 Bonisteel Boulevard, Ann Arbor, Michigan 48109, USA}

\author{Yang Zhang}%
\affiliation{Department of Astrophysical Sciences, Princeton University, Princeton, New Jersey 08544, USA}
 \affiliation{University Corporation for Atmospheric Research, Boulder, CO, 80301, USA 
}

\author{Stepan S. Bulanov}
\affiliation{Lawrence Berkeley National Laboratory, Berkeley, California 94720, USA}

\author{Sergei V. Bulanov}
\affiliation{ELI Beamlines Facility, The Extreme Light Infrastructure ERIC, Za Radnic\'{i} 835, Doln\'{i} B$\check{r}$e$\check{z}$any, 25241, Czech Republic}
\affiliation{National Institutes for Quantum and Radiological Science and Technology (QST),
Kansai Photon Science Institute, Kyoto, 619-0215 Japan
}

\author{Hui Chen}%
\affiliation{Lawrence Livermore National Laboratory, Livermore, California 94550, USA}

\author{Lan Gao}%
\affiliation{Princeton Plasma Physics Laboratory, Princeton University, 100 Stellarator Rd, Princeton, NJ, 08540, USA}

\author{Gabriele M. Grittani}
\affiliation{ELI Beamlines Facility, The Extreme Light Infrastructure ERIC, Za Radnic\'{i} 835, Doln\'{i} B$\check{r}$e$\check{z}$any, 25241, Czech Republic}

\author{Xiaocan Li}
\affiliation{Los Alamos National Laboratory, Los Alamos, New Mexico 87545, USA}

\author{Kian Orr}%
\affiliation{Department of Mechanical and Aerospace Engineering, Princeton University, Princeton, New Jersey 08544, USA}

\author{Geoffrey Pomraning}%
\affiliation{Department of Astrophysical Sciences, Princeton University, Princeton, New Jersey 08544, USA}

\author{Kevin M. Schoeffler}%
\affiliation{Institut für Theoretische Physik, Ruhr-Universität Bochum, Bochum 44801, Germany}

\author{Alexander G. R. Thomas}
\affiliation{G\'{e}rard Mourou Center for Ultrafast Optical Science, University of Michigan, 2200 Bonisteel Boulevard, Ann Arbor, Michigan 48109, USA}

\author{Hantao Ji}%
\affiliation{Department of Astrophysical Sciences, Princeton University, Princeton, New Jersey 08544, USA}
\affiliation{Princeton Plasma Physics Laboratory, Princeton University, 100 Stellarator Rd, Princeton, NJ, 08540, USA}

\date{\today}

\begin{abstract}

We propose and simulate a laboratory platform to study the effects of positrons in magnetic reconnection using laser-driven capacitor coils. Using particle-in-cell simulations, we show that externally injected MeV electron-positron pairs are trapped in the coil current sheet, significantly modifying the reconnection dynamics and particle acceleration. These pairs increase the reconnection rate by a factor of approximately 8, which Ohm's law decomposition reveals to be driven by the divergence of the generalized pressure tensor. Based on their high energy and magnetization, the pairs also substantially broaden the diffusion region. Particle tracking simulations in realistic coil magnetic fields further demonstrate that injected pairs can remain confined for several picoseconds, providing conditions for sustained interaction with the reconnection region. These results establish a near-term pathway to laboratory studies of positron-influenced reconnection, bridging high-energy-density experiments with pair-dominated astrophysical environments.

 \end{abstract}

\maketitle

\begin{figure}
    \includegraphics{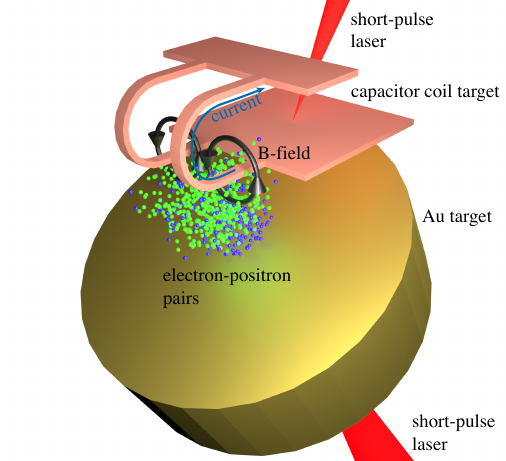}
    \caption{Proposed experimental concept where pairs from a laser-irradiated Au foil are injected into the coils of a laser-driven capacitor coil. Magnetic fields produced around the legs of the coil create a reconnection configuration. The sub-millimeter separation between the legs allows for electron-only magnetic reconnection with pair loading. Injected pairs are represented by green and blue spheres.  
    }
    \label{fig:Capcoilconcept}
\end{figure}

Magnetic reconnection is a widely studied process in both the astrophysics and laboratory astrophysics communities \cite{Yamada_RevModPhys_2010, ji22}. This ubiquity comes largely from the numerous environments in our universe where magnetic reconnection is believed to occur and the demonstrated success of laboratory experiments in creating conditions relevant to these environments \cite{stenzel79,Nilson_PRL_2006,olson16,Hare_PRL_2017,shi22,zhang2023generation}. Magnetic reconnection, the process whereby magnetic field energy is rapidly converted to particle kinetic energy, has been invoked in various systems as the cause of highly energetic phenomena, e.g. solar flares \cite{Forbes_JGR_2000}, and may be responsible for high energy particle generation from the extreme environments surrounding magnetars \cite{Uzdensky_SpaceScieRev_2011}. 

The theoretical studies of magnetic reconnection cover a large parameter space of field strengths, plasma densities, and plasma composition. These properties can greatly affect the rate of reconnection and the form of acceleration that may occur \cite{Uzdensky_2022}. Laboratory experiments similarly study a wide range of parameters \cite{Nilson_PRL_2006,Raymond_PRE_2018}; however, there exists a significant discrepancy in the plasma composition considered in theory and experiment, specifically in the extreme astrophysics community. Due to the energetic conditions in the extreme environments considered by this community, e.g., surrounding black holes, pulsars, and active galactic nuclei (AGN) jets there should be significant electron-positron pair creation \cite{Goldreich_APJ_1969,Timokhin_2015}, therefore reconnection happening in these environments should be partially or fully composed of pairs \cite{Ripperda_ApJ_2022,Hakobyan_2023}. The dynamics of pair reconnection or mixed pair-ion reconnection can greatly differ from electron-ion reconnection \cite{Werner_MNRAS_2018,Schoeffler_ApJ_2019,Petropoulou_2019,Chernoglazov_2023,Figueiredo_AA_2024,Imbrogno_2025}.

There is significant interest within the laser-plasma community in producing sufficient pairs to generate plasmas to study plasma processes. These efforts focused on a few experimental schemes for pair production based on the Bethe-Heitler process as recently reviewed by Chen and Fiuza \cite{Chen_PoP_2023}. In the most commonly used ``direct'' method, a laser incident on a high-Z target accelerates electrons. These electrons generate high energy photons through bremsstrahlung that can be of sufficient energy to generate pairs when interacting with the Coulomb fields of the high-Z target. Experimentally, this process has generated $\sim10^{12}$ pairs using relativistic intensity kJ-class lasers \cite{Jiang_APL_2021}. Magnetic mirror schemes have been used to trap the generated pairs for future basic plasma studies \cite{vonderLinden_PoP_2021}. However, the use of laser-generated pairs has been limited \cite{Warwick_PRL_2017}. 

While the study of plasma processes with pairs has generally been viewed to be limited due to the pair density requirements, here we consider a pair-loaded electron-only reconnection platform that may be implemented using currently available lasers. Theory and simulation are used to study the injection of pairs into a laser-driven magnetic reconnection platform and the reconnection physics that may be enabled by this platform. We build from the capacitor coil platform \cite{ji24} that has provided important insight into reconnection dynamics \cite{Zhang_NatPhys_2023} and particle acceleration \cite{Chien_NatPhys_2023}. We find that pairs may be injected into the coils and confined over several picoseconds based on magnetic field strengths from short-pulse driven coils \cite{gao_2025}. Depending on the density of injected pairs, they may either act as a tracer of the reconnection process or significantly modify the reconnection dynamics and energy gained. The experimental implementation of this concept will significantly reduce the expected timeline for reconnection studies including positrons, providing insight into reconnection as it happens in the most extreme environments in the universe using currently available kJ laser systems. Furthermore, the method of injection can be interpreted as an analog for pair injection in reconnection with strong magnetic fields where the effects of pair creation has seen significant interest \cite{Beloborodov_2017,Mehlhaff_MNRAS_2023}.  

Figure~\ref{fig:Capcoilconcept} demonstrates the concept for the experiment designed around a laser-driven capacitor coil. The capacitor coil target consists of two $1.5\times1.5$ mm foils connected by two wires separated by $\sim600\;\mu$m. The back plate is irradiated by a laser pulse, heating electrons and driving the formation of a plasma plume and an electric potential between the plates. This potential results in a return current that runs through the wires, generating magnetic fields. These azimuthal magnetic fields are anti-parallel in the region between the wires and can reconnect. Recent results using a 15 ps duration, intense $I_L\sim10^{19}$ W/cm$^{2}$ laser pulse demonstrated a current of 120 kA and a few hundred Tesla magnetic fields \cite{gao_2025}. A thick Au foil is irradiated by a second high-intensity kJ laser pulse, generating a high-divergence beam with a duration similar to that of the laser pulse \cite{Chen_PoP_2023}. The target angle and position must be chosen to maximize injection and confinement of pairs as will be discussed later in this Letter. 

\begin{figure*}
    \includegraphics{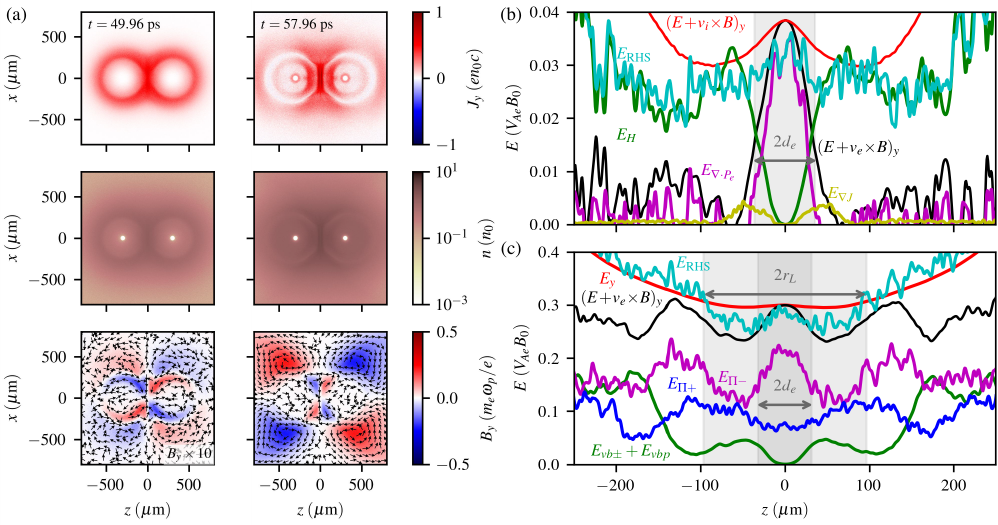}
    \caption{Two-dimensional VPIC simulation of reconnection with pair injection with a density $\sim$30\% of the background electron-ion plasma in the current sheet. The evolution of the out-of-plane current density $J_y$, electron density and out-of-plane magnetic field $B_y$ are shown in (a) where 1 MeV pairs are injected from 50-60 ps. In row 3, arrows show the direction of the in-plane current. Plots (b) and (c) demonstrate the decomposition of the reconnection electric field into the terms given by Ohm's law at $t = 57.96$ ps for a simulation without and with the injection of pairs respectively. The electric field has been normalized by the electron Alfv\'{e}n speed $V_{Ae}$ multiplied the magnetic field $B_0$ calculated at $z= 1d_e$. This region has been marked in gray in (b) and (c). See Appendix A for the definition of the Ohm's law terms that are labeled in (b) and (c) with colors matching the corresponding lines.
    }
    \label{fig:pairevo}
\end{figure*}

\begin{figure}
    \includegraphics{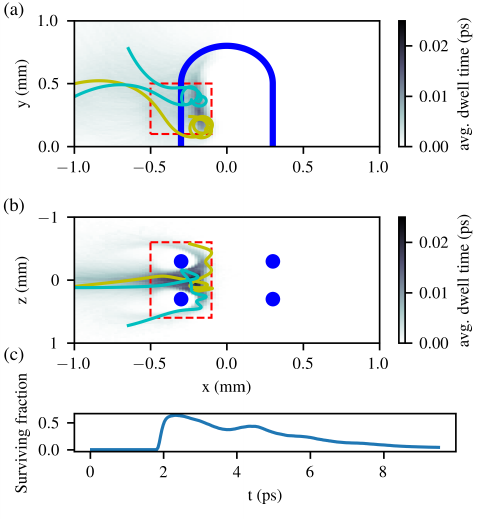}
    \caption{Trapping of positrons between the coils using the magnetic fields 50 ps into the 100 ps current rise time. Plots (a) and (b) show the integrated time spent by pairs in the x-y and x-z planes respectively. Trapped trajectories of positrons are shown with overlays for the coil positions in blue. The fraction of particles within the red rectangle in (a) and (b) are plotted in (c) showing trapping over several ps. 
    }
    \label{fig:trapping}
\end{figure}

Without the inclusion of pairs, the capacitor coil drives electron-only push-type reconnection \cite{Chien_PoP_2019}. The coils are ionized by ohmic heating and x-rays from the laser interaction on the backplate, forming a plasma. Growing magnetic fields are simultaneously generated around the coils that expand with magnetized electrons into the mid-plane where reconnection occurs. In nanosecond long-pulse laser-driven reconnection, the measured electron density is $n_e\sim10^{18}$ cm$^{-3}$ \cite{Zhang_NatPhys_2023} and is similarly dense on the nanosecond timescale for short pulse interactions \cite{zhang2026}. The distance between the coils is approximately one ion skin-depth and previous experiments have observed electron acceleration consistent with electron-only reconnection \cite{Chien_NatPhys_2023}, a regime relevant to the Earth's magnetosheath \cite{Phan_Nature_2018_e_only_magnetosheath}. 

Let us first consider the necessary conditions for pairs to support reconnection alone. They must have a sufficient density to sustain a current sheet and must exist over a large enough volume to support the current sheet and allow reconnection. The minimum density can be estimated from Amperes law using a current sheet of thickness $\delta$ across which the B-field varies due to a current density $j = B/\mu_0\delta$. The current density can be formed by relativistic pairs drifting in opposite directions approximately at the speed of light. If the density is large, the current sheet width will become very small; however, it is limited to be larger than the electron skin depth $d_e = c/\omega_p$, where $\omega_p$ is the plasma frequency. For the pairs to be magnetized and reconnect, the separation between the coils $L_{sep}$ must be larger than the Larmor radius $r_L = p_\perp/eB$, where $p_\perp$ is the momentum perpendicular to the magnetic field. With a separation between the coils of $L_{sep}=600\; \mu$m and a 120 kA current as demonstrated experimentally by Gao \textit{et al.} \cite{gao_2025} a magnetic field of 100 T at $150\; \mu$m from one of the coils is expected. Using the conditions that both the minimum current sheet width and $r_L$ are both smaller than $L_{sep}/4$ predicts that reconnection should be possible for few MeV pairs ($\gamma < 10$) with densities $>10^{16}$ cm$^{-3}$. Note, the fields used here are near the peak fields generated by the coil. It will therefore be necessary to inject the pairs just before the coil fields have reached their peak strength such that the pairs are trapped, but will still be driven into the mid-plane as the fields continue to increase in strength. 

To understand how this density of MeV pairs can be achieved we can consider the properties of experimentally demonstrated beams \cite{Chen_PoP_2023}. The pairs with properties most similar to what is required here were generated in the interaction of high energy laser pulses with thick Au targets. These beams generated either at OMEGA EP \cite{Chen_PRL_2015} or the Texas Petawatt \cite{Liang_SR_2015} had durations of a few ps, energies $>5$ MeV, positron to electron ratios $>0.2$, and positron numbers $10^{10} -10^{12}$. The divergence ($<50 \degree$ full-width half-maximum) and spot size of the pair beams was large which limited the density. However, in the case of the capacitor coil, injected pairs can travel freely along magnetic field lines and fill the area around the coils. For coils with $L_{sep}=600\;\mu$m, let us assume that the area of the filled will be $\sim 1\times 1\;$mm$^2$. A lower bound on the layer thickness is the Larmor radius, which is $\sim20\; \mu$m for 1-MeV pairs. To achieve a density of $10^{16}$ cm$^{-3}$ this requires at least $2\times 10^{11}$ pairs. This is just within the bounds of what has been demonstrated experimentally. However, due to the time variance of the fields and imperfect trapping, it is unlikely that reconnection can be supported by pairs alone. Instead, a more realistic experiment using currently available lasers is to introduce pairs into a current sheet that has already formed. However, the multi-MeV pairs that have been demonstrated experimentally have energies slightly greater than what can be trapped by the fields of the capacitor coil. It will therefore be necessary to optimize for the generation of MeV pairs instead of multi-MeV pairs which can be done by increasing the radius of the Au target \cite{Chen_PRL_2010}.

To understand how electron-only reconnection is affected by the injection of pairs, 2D simulations were run using the particle-in-cell code VPIC \cite{Bowers_PoP_2008}. Currents were injected in circular regions with a radius of $25\;\mu$m and separated by $600\;\mu$m with a linear rise over 100 ps to a peak current of 120 kA. While the rise time for the current in these coils has not been measured, recent measurements under comparable conditions have shown similar rise times \cite{Zhang_PoP_2025}. A neutral electron-ion plasma was injected with a reference density $n_0 =10^{16}$ cm$^{-3}$ from the coils as would be expected from short-pulse driven coils for these early times. The ions had an ionization state $Z = 20$ and a normalized mass 63.55 times the mass of a proton to represent a partially ionized Cu ion. A uniform density of $10^{15}$ cm$^{-3}$ was initialized throughout the box which was found to reduce growing fields that affect numerical stability at the boundaries. The electrons and ions in the initial and injected populations had a temperature of 10 keV. At this temperature the chosen cell size $\Delta x=\Delta z = 0.1L_{D} = r_L/3 = d_e/71$, where $L_D$ is the Debye length, $r_L$ is the Larmor radius, and $d_e$ is the electron skin depth. All particles were injected or initialized with 100 particles-per-cell which was found to be sufficient to resolve the reconnection physics and particle acceleration. Absorbing boundaries were used for particles and fields. A large box with dimensions of 3 mm $\times$ 3 mm was used to remove boundary effects from the reconnection dynamics. While the electron-ion plasma was injected on a linear ramp, the electron-positron pairs were injected as a mono-energetic 1 MeV population with a constant injected density from 50-60 ps. The initial direction of propagation for each particle in the population was sampled from a random uniform distribution. This was done to create a simplified version of the beam as it would be injected from a 10 ps laser pulse interacting with an Au foil and trapped between the coils. Motivation for this injected population will be demonstrated later in this Letter. 

Figure \ref{fig:pairevo}(a) shows the evolution of a simulation where the density of the pairs is approximately 30\% of the background electron density in the current sheet. Prior to the pairs being injected, this system evolves and reconnects. With the chosen initial and injected density, sufficient time is required for the density to build up and support the current sheet. The pairs are injected at 50 ps, a time when reconnection is already occurring, however their arrival sparks several changes in the reconnection dynamics. As the pairs are injected they greatly increase the reconnection rate $R_{EDR}=E_y/V_{Ae}B_0$, calculated using local values at $z=1d_e$, i.e., in the upstream of the electron diffusion region (EDR). This is the method reported by Liu \textit{et al.} for electron-only reconnection \cite{Liu2025_eonly}, but with a relativistic Alfv\'en speed $V_A = c \sqrt{\sigma/(1 + \sigma)}$ where $\sigma = \sigma_{cold,e} = B^2/\mu_0(n_e+n_p)m_ec^2$ is the magnetization of the electrons with density $n_e$ and positrons with density $n_p$. In Fig.~\ref{fig:pairevo}(b) and (c) the $y$-directed reconnection electric field is shown for a simulation without and with the injection of pairs respectively. The E-field has been time-averaged over 0.6 ps, spatially averaged over the interval $x = [-7.1,7.1]\; \mu$m, and normalized by $V_{Ae}B_0$ taken at $d_e$ which is marked by a gray region in both plots. We observe a factor of approximately 8 increase in the reconnection rate with the injection of pairs. 

To understand the increase in the reconnection rate we decompose the E-field into the terms given by the electron-ion Ohm's law in Fig.~\ref{fig:pairevo}(b) \cite{Liu2025_ohms}, and the relativistic electron-positron-ion Ohm's law in (c) \cite{Imbrogno_2025} (see Appendix A for details on these equations). In Fig.~\ref{fig:pairevo}(b) we observe that the electron pressure term $E_{\nabla \cdot P_e}$ dominates within the EDR and the Hall term $E_H\propto (\mathbf{J}\times\mathbf{B})_y$ dominates outside the EDR. In addition to the term $E_{\nabla J}\propto (\mathbf{J}\cdot\nabla(\mathbf{J}/n_e))_y$ these terms make up the right-hand-side (RHS) of Ohm's law $E_{RHS}$ which is close to the left-hand-side (LHS) $E_{LHS} = E_y + (\mathbf{v_i}\times \mathbf{B})_y$. We have neglected $E_{\partial_t}$ which is small and affected by the injection of particles at the coils. Outside a few $d_e$, the approximation that the electron velocity greatly exceeds the ion velocity ($\mathbf{v_e}\gg \mathbf{v_i}$) breaks down and the sides of Ohm's law do not agree. In Fig.~\ref{fig:pairevo}(a) row 3 we observe a significant increase in the quadrupolar magnetic field, therefore it may be expected that the Hall term would explain the increase in the reconnection rate, however this is not the case. In Fig.~\ref{fig:pairevo}(c) $E_{vb\pm}+E_{vbp}$, which is conceptually similar to the electron-ion Hall term, shows a similar amplitude to the Hall term in the case without injection. Instead, it is the terms $E_{\Pi +}$ and $E_{\Pi-}$, coming from the divergence of the off-diagonal elements of the generalized pressure tensor for the positrons and electrons respectively, that dominate the electric field. Additionally, while $d_e$ is similar for both cases, $(\mathbf{E} + \mathbf{v}_e\times \mathbf{B})_y$ is significantly broadened and gains two peaks with the injection of pairs. Due to their much higher energy, the pairs have an effectively larger diffusion region given approximately by the region where the Larmor radius is greater than the distance to the x-point, i.e., $r_L(z) \geq z$ with the x-point at $z=0$. This region has been marked in light gray in Fig.~\ref{fig:pairevo}(c). In summary, when high energy pairs are injected they increase the reconnection rate by increasing the plasma pressure and adding a diffusion region that is defined by the magnetization of the injected pairs. 

Notably, this system has a $\sigma_{cold,e}\sim 0.16$ at $z = 1d_e$ and exceeds unity further into the upstream, indicating that the inflowing plasma transitions into the relativistic regime for electron-only reconnection. It is therefore expected that the particles should gain a significant amount of energy exceeding their rest mass. The pairs start at 1 MeV and after less than 10 ps the injected electrons gain 1.8 MeV, the background electrons gain 1.9 MeV, and the injected positrons 2.2 MeV. Injecting with only a single coil (no reconnection) gives an energy gain of 1.6 MeV for injected positrons and electrons and 1.4 MeV for background electrons. With 2 coils and no injection the background electrons only gain 0.8 MeV. While this shows that the pairs are energized by reconnection, measuring this change in energy may be masked if experimental noise is significant. Furthermore, the effects of the 3D reconnection geometry on the acceleration of these particles should be understood, however this is outside the scope of this study. 

Finally, we must consider how the pairs will be injected into the coils and how they can be sustained for the several picoseconds necessary for acceleration. We calculate the magnetic fields of the coils using the Biot-Savart law and propagate particles in 3D through these fields using the relativistic Boris pusher implemented in the code PlasmaPy \cite{plasmapy2026}. While electric fields are generated from the time varying magnetic field, these fields do not appear to be significant from our experimental results \cite{Zhang_PPCF_2025}. A beam consisting of $10^4$ positrons was injected at $(x,y,z)=(-1,0.4,0)$ mm with an angle of $30\degree$ from the $x$-axis towards the $y$-axis. The beam had a Gaussian energy spread $\sigma_E/E_0=10\%$, a Gaussian distribution in solid angle with a $1/e$ half-angle divergence $\theta_0=15\degree$, and a Gaussian transverse profile with $1/e$ radius $r_0=250\;\mu$m. Figure \ref{fig:trapping} shows the amount of time that these particles stay in each part of the simulation domain integrated over the $z$-direction in (a), the $y$-direction in (b), and divided by the total number of particles. Particles bounce between the coil fields and perform several orbits as shown. However, many particles do not follow trapped trajectories and quickly leave the coils. To quantify this, the fraction of particles within the volume marked in red in Fig.~\ref{fig:trapping}(a-b) was calculated and plotted in (c). Almost 60\% of particles make it into this region with approximately 50\% of those lasting more than 4 ps. For a real pair beam with a duration of $\sim10$ ps, this means that particles from the front of the beam will continue to exist between the coils as more pairs are injected, increasing the density. The chosen angle for injection was found by integrating the surviving fraction over time for several beam angles. We find that the injection is robust, showing similar values for the integrated surviving fraction for $\pm10\degree$ from this position. However, it should be noted that the angle to trap electrons and positrons is not the same, likely resulting in the injection of unequal numbers of electrons and positrons. 

In conclusion, we have proposed a magnetic reconnection experiment to study the effects of positrons. This experiment is based on the well-tested capacitor coil platform that was recently driven by relativistic-intensity short pulse lasers \cite{gao_2025}. We have demonstrated that pairs from laser-driven Au foils are readily injected into these coils, can exist in the current sheet for several ps, and significantly modify the reconnection dynamics. While the simulations performed here were limited to 2D due to computational cost, they show significant modification to the reconnecting fields, dynamics, and the accelerated particles. Experimentally, these can be diagnosed using electron-positron spectrometers and proton radiography. These results demonstrate that positron-influenced reconnection can be studied with existing kJ-class lasers, bridging the gap between laboratory experiments and pair-dominated astrophysical plasmas for the first time.

We thank Dr. Hayk Hakobyan for useful discussions. This work was supported by the US Department of Energy High-Energy-Density Laboratory Plasma Science program under Grant No. DE-SC0020103 and by the NSF under grant No. 2512021. Q.Q. and A.G.R.T were supported by NSF grants 2108075, 2126181, and 2206059. R.F. and A.G.R.T were supported by US DOE National Nuclear Security Administration (NNSA) Center of Excellence under Cooperative Agreement No. DE- NA0003869. Y.Z. was supported by the NASA Living with a Star Jack Eddy Postdoctoral Fellowship Program, administered by UCAR's Cooperative Programs for the Advancement of Earth System Science (CPAESS) under award $\#$80NSSC22M0097. This research was supported in part by grant no. NSF PHY-2309135 to the Kavli Institute for Theoretical Physics (KITP).

This report was prepared as an account of work sponsored by an agency of the United States Government. Neither the United States Government nor any agency thereof, nor any of their employees, makes any warranty, express or implied, or assumes any legal liability or responsibility for the accuracy, completeness, or usefulness of any information, apparatus, product, or process disclosed, or represents that its use would not infringe privately owned rights. Reference herein to any specific commercial product, process, or service by trade name, trademark, manufacturer, or otherwise does not necessarily constitute or imply its endorsement, recommendation, or favoring by the United States Government or any agency thereof. The views and opinions of authors expressed herein do not necessarily state or reflect those of the United States Government or any agency thereof.

The data that support the findings of this study are available from the corresponding author upon reasonable request.

\bibliography{compressed_bib}

\section*{End Matter}

\renewcommand{\theequation}{A\arabic{equation}}
\setcounter{equation}{0}
\textit{Appendix A: Ohm's law calculation}---In this Letter we apply two versions of Ohm's law: one to decompose the reconnection electric field in an electron-ion plasma, and another for an electron-positron-ion plasma. The former can be found in the review by Liu \textit{et al.} \cite{Liu2025_ohms} and the latter in the recent work of Imbrogno \textit{et al.} \cite{Imbrogno_2025}. The electron-ion Ohm's law is derived from the electron momentum equation in a collisionless plasma assuming the electron velocity $\mathbf{v_e}$ is much larger than the ion velocity $\mathbf{v_i}$ such that the current $\mathbf{J}\approx -en_e\mathbf{v_e}$. It is written in SI units as \cite{Liu2025_ohms}, 
\begin{multline}\label{Eqn:EI_Ohms}
        \mathbf{E} + \mathbf{v_i\times B}=\frac{\mathbf{J\times B}}{n_ee} - \frac{\nabla\cdot \mathbf{P_e}}{n_ee} \\ - \frac{m_e}{e^2}\left(\frac{\mathbf{J}}{n_ee}\right)\cdot\nabla\left(\frac{\mathbf{J}}{n_e}\right) + \frac{m_e}{e^2}\frac{\partial}{\partial t}\left(\frac{\mathbf{J}}{n_e} \right) \\
        = \mathbf{E}_H + \mathbf{E}_{\nabla\cdot P_e} + \mathbf{E}_{\nabla J} + \mathbf{E}_{\partial_t},
\end{multline}
where $\mathbf{P_e}$ is the electron pressure tensor, $n_e$ is the electron density, and $m_e$ and $e$ are the electron mass and charge respectively. These terms are plotted in Fig.~\ref{fig:pairevo}(b). On the right-hand-side (RHS) of the equation we find, consistent with previous work, that the first term dominates just outside the EDR, while the second term dominates within the EDR. The third term adds a small contribution, while the final term is found to be small and negative in our simulations. The injection rate directly contributes to this term, making it sensitive to the coil driving conditions rather than the reconnection dynamics alone. It has therefore been neglected from Fig.~\ref{fig:pairevo}(b). Finally, we observe a discrepancy between the LHS and RHS of the equation in our calculations outside the EDR. In this region the $\mathbf{v_e}\gg \mathbf{v_i}$ assumption does not hold as the electron to ion velocity ratio is only approximately 2.5. 

The electron-positron-ion Ohm's law was derived relativistically and is given by \cite{Imbrogno_2025},
\begin{multline}
    \mathbf{E} = \frac{m_-}{e^2\mathcal{N}}\frac{\partial\mathcal{J}}{\partial t} - \frac{1}{\mathcal{N}}(n_+\mathbf{V}_+ + n_-\mathbf{V}_-)\times \mathbf{B} \\ - \frac{1}{\mathcal{N}}n_p\mathbf{V}_p\times \mathbf{B} - \frac{1}{e\mathcal{N}}\nabla\cdot\mathbf{\Pi}_- \\ + \frac{1}{e\mathcal{N}}\nabla\cdot\mathbf{\Pi}_+ + \frac{m_-}{m_p}\frac{1}{e\mathcal{N}}\nabla\cdot\mathbf{\Pi}_p \\ = \mathbf{E}_{\partial_t} + \mathbf{E}_{vb\pm} + \mathbf{E}_{vbp} + \mathbf{E}_{\Pi-} + \mathbf{E}_{\Pi+} + \mathbf{E}_{\Pi p}.
\end{multline}
Here, the electrons, positrons, and ions are denoted by the subscript $a = -,+,$ and $p$ respectively. The total current density is given by $\mathbf{\mathcal{J}} = \Sigma_a q_an_a\mathbf{U_a}$, where $\mathbf{U}_a = \gamma \mathbf{V}_a$ is the bulk four-velocity with $\mathbf{V}_a$ as the three-velocity and $\gamma$ the Lorentz factor. The effective number density $\mathcal{N} = n_pm_-/m_p + n_+ + n_-$ where $m_a$ is the mass and $n_a$ is the density of each species. The generalized pressure tensor $\mathbf{\Pi}_a = \mathbf{P}_a + m_an_a\mathbf{V}_a\mathbf{U}_a$ includes the bulk contributions that are excluded from $\mathbf{P}_e$ in Eq.~\eqref{Eqn:EI_Ohms}. In our analysis we neglected the $\mathbf{E}_{\partial_t}$ term which is unphysical as pairs are injected continuously near the x-point breaking the conservation of momentum. We also neglected $\mathbf{E}_{\Pi p}$ which is much smaller than the other terms for our conditions.  

\end{document}